\documentstyle[twoside,fleqn,espcrc2]{article}
\input{epsf}
\newcommand{\AmS}{{\protect\the\textfont2
  A\kern-.1667em\lower.5ex\hbox{M}\kern-.125emS}}

\title{Intrinsic Geometric Structure of $c=-2$ Quantum Gravity}

\author{J.Ambj\o rn\address{Niels Bohr Institute, Blegdamsvej 17 \\
2100 Copenhagen \O , Denmark} 
       , K.N.Anagnostopoulos$^{{\rm a}}$, T.Ichihara\address{Tokyo Institute of 
Technology, O-okayama \\ Meguro, Tokyo, Japan}
       , {\bf L.Jensen}$^{{\rm a}}$, N.Kawamoto\address{Department of Physics, Hokkaido 
University \\ Sapporo, Japan}
       , Y.Watabiki$^{{\rm b}}$ and K.Yotsuji$^{{\rm c}}$ }

\begin{document}

\begin{abstract}

We couple $c=-2$ matter to 2-dimensional gravity within the framework of
dynamical triangulations. We use a very fast algorithm, special to the
$c=-2$ case, in order to test scaling of correlation functions defined
in terms of geodesic distance and we determine the fractal dimension
$d_H$ with high accuracy. We find $d_H=3.58(4)$, consistent with a
prediction coming from the study of diffusion in the context of
Liouville theory, and that the quantum space--time possesses the same
fractal properties at all distance scales similarly to the case of pure
gravity.

\end{abstract}

\maketitle

\section{The model}

Our starting point is the partition function of Euclidean 2d quantum
gravity at fixed volume and discretize it by using dynamical
triangulations with a fixed number of equilateral triangles $N$. The
integral over metrics changes into a sum over triangulations ${\cal
T}_N$, and by coupling the quantum gravity to $c$ Gaussian fields, the
partition function is rewritten as

\begin{equation}
Z_{N} = \sum_{{\cal T}_{N}} \frac{1}{{\cal S}_{{\cal T}_{N}}}(det'{\cal C}_{{\cal T}_{N}})^{-c/2}
\end{equation}

Here ${\cal C}_{T_{N}}$ is the adjacency matrix of the $\phi^{3}$
graph, which is dual to ${\cal T}_{N}$, ${\cal S}_{{\cal T}_{N}}$ is a
symmetry factor, and $det'{\cal C}_{{\cal T}_{N}}$ denotes the
determinant of ${\cal C}_{{\cal T}_{N}}$ with the zero-eigenvalue
removed.  This partition function also serves as the defining equation
for coupling to matter with $c<0$. We restrict ourselves to surfaces
with the topology of a sphere, for reasons which will soon be clear.
By selecting $c=-2$ it is possible to construct a recursive sampling
algorithm for the numerical generation of an ensemble of surfaces
\cite{kkm}. Two results from graph theory are crucial here.  First,
the determinant (appropriately defined) of ${\cal C}_{T_{N}}$ is equal
to the number of spanning trees in the $\phi^{3}$ graph. Second, a
graph can be embedded on a sphere if - and only if - it is planar.
Thus restricting ourselves to configurations with the topology of a
sphere, we can decompose the sum over all triangulations into a sum
over all combinations of $\phi^{3}$ trees and rainbows - a rainbow
being a set of non-crossing lines which connect the endpoints of the
tree-graphs. Both the trees and the rainbows satisfy Schwinger-Dyson
equations, and by solving these equations, we can generate the correct
ensemble of configurations using a recursive sampling algorithm. The
time for generating a configuration grows linearly with its size, a
rare case of a non--trivial statistical system with subexponential
time growth of statistically completely independent
configurations. This fact has enabled us to generate a substantial
number of configurations with up to 8 million triangles.

\section{Definition of the fractal dimension}

The fractal dimension is defined from scaling of correlation functions
defined in terms of geodesic distance $r$, therefore we have to define
geodesic distance on the triangulated surfaces. This is done in two
ways. Either as the number of links between vertices in the $\phi^{3}$
graph (the dual lattice), or the number of links between the vertices
of the triangles (the direct lattice). It is a non--trivial fact that
both definitions are proportional to each other after taking the
quantum average.  In order to study the intrinsic structure, consider
the average volume $n_{N}(R)$ of a spherical shell at a distance R.
We can then define {\it two} fractal dimensions $d_{h}$ and $d_H$ in
the following way:
\begin{equation}
n_{N}(r) = N^{1-1/d_{H}}F_{1}(x), \; \; \; x = \frac{r+a}{N^{1/d_{H}}}.
\end{equation}
where $F_1(x)\sim x^{d_h-1}$ for $x\sim 0$. The existence of the
scaling variable $x$ at all length scales is a non--trivial fact to be
tested by the simulations. The ``shift'' $a$ is a finite size
correction. If space--time looks the same at all scales then
$d_h=d_H$.  Our simulations provide evidence, that this is indeed the
case.  For the correct value of $d_{H}$ and $a$, the distributions
$n_{N}(r)$, plotted as a function of $x$, will fall on top of each
other if they actually do scale.  That this is indeed the case, and
that the finite-size correction $a$ is necessary, can be seen in
fig. 1 and fig. 2 were the dual lattice distance has been used and the
configuration sizes range from 2k to 8M triangles.

\begin{table*}[hbt]
\setlength{\tabcolsep}{1.5pc}
\newlength{\digitwidth} \settowidth{\digitwidth}{\rm 0}
\catcode`?=\active \def?{\kern\digitwidth}
\caption{The fractal dimension determined from different methods.}
\label{tab:effluents}
\begin{tabular*}{\textwidth}{llll}
\hline

  & & lattice & method \\
\hline
  $d_{h}$ & $3.58(5)$ & direct & Short distance scaling \\
  $d_{h}$ & $3.50(4)$ & dual & Short distance scaling \\
  $d_{H}$ & $3.60(3)$ & direct & Collapsing the 2-point function \\
  $d_{H}$ & $3.55(3)$ & dual & Collapsing the 2-point function \\
  $d_{H}$ & $3.577(8)$ & direct & Intersection of $R_{a,N}$ \\
  $d_{H}$ & $3.574(20)$ & dual & Intersection of $R_{a,N}$ \\
  $d_{H}$ & $3.64(7)$ & direct & Collapsing of higher moments \\

\hline
\end{tabular*}
\end{table*}

\begin{figure}[htb]
\vspace{5mm}
\epsfxsize7cm
\epsfysize5.5cm
\mbox{\epsffile{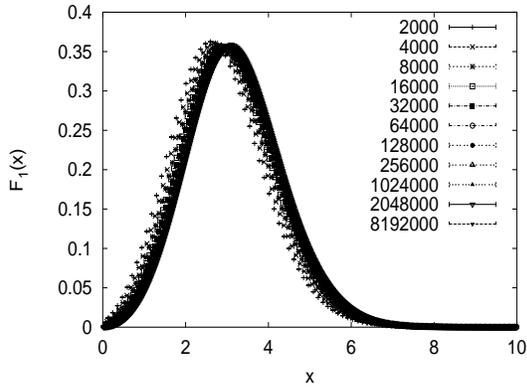}}
\vspace{-1cm}
\caption{$F_{1}(x)$ without shift $(a=0)$}
\label{fig:ntdH3.56a0.00.eps}
\end{figure}
\begin{figure}[htb]
\vspace{2mm}
\epsfxsize7cm
\epsfysize5.5cm
\mbox{\epsffile{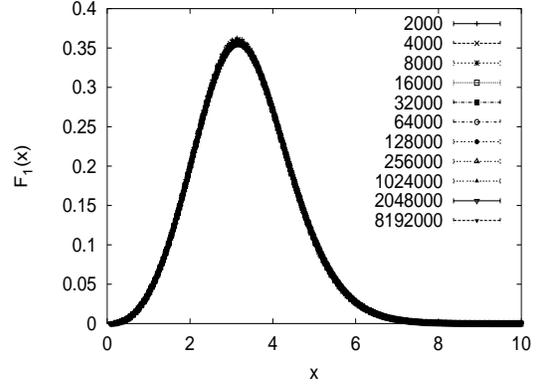}}
\vspace{-1cm}
\caption{$F_{1}(x)$ with shift $(a=4.50)$}
\label{fig:nxtdH3.56a4.50.eps}
\vspace{-1.2cm}
\end{figure}

\begin{figure}[htb]
\vspace{0mm}
\epsfxsize7cm
\epsfysize5.5cm
\mbox{\epsffile{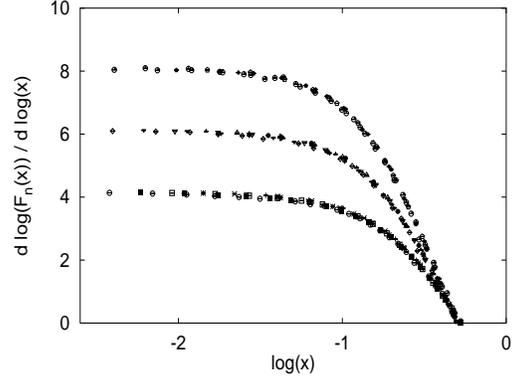}}
\vspace{-1cm}
\caption{Short distance higher moments}
\label{fig:shortdist}
\end{figure}

\section{Higher moments}

We can also consider the higher moments {$\langle l^{n}\rangle_{r,N}
$} of the boundary length $l$. These are defined in terms of the
loop--length distribution function $\rho(r,l)$, which counts the
number of disconnected loops of the set of points at distance $r$ from
a given point, as $\langle l^{n}\rangle_{r,N}= \int
dl\,l^n\,\rho(r,l)$.  From Liouville theory we expect $[l^{2}] = [A]$,
but the simulations are consistent with the scaling
\begin{equation}
\label{*ll}
\langle l^{n}\rangle_{r,N}=N^{\frac{2n}{d_H}}F_n(x)
\; \; \; \; , n\geq 2\, ,
\end{equation}
where $F_n(x)\sim x^{2n}$ for $x\ll 1$.
This implies that $[l^{d_{H}/2}] = [A]$. 
The small distance scaling can be seen clearly in fig.3 where we show
a plot of the logarithmic derivative of the moments $n=2,3,4$.
Trying to collapse the distributions of the moments 
is consistent with eq.(\ref{*ll}).

\section{Determination of fractal Dimension}

Using different methods to collapse the distributions \cite{our}
we can determine both $d_{H}$ and $d_{h}$ with great precision. The
results using these methods are shown in table 1. 
$R_{a,N}(d_H)=1/N^{1+1/dH}\sum_{r}(r+a)n_N(r)$ is the average radius of the
universe \cite{our} and $d_H$ can be determined by tuning $d_H$ and
$a$ so that the value of $R_{a,N}(d_H)$ is independent of $N$. As is shown
in fig.4 with the link distance, this works very well indeed.
\begin{figure}[htb]
\vspace{-5mm}
\epsfxsize7cm
\epsfysize5.5cm
\mbox{\epsffile{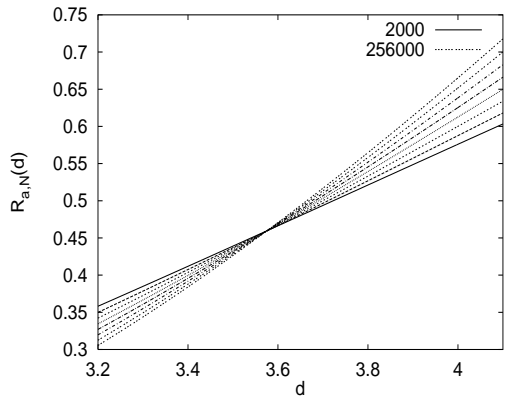}}
\vspace{-1cm}
\caption{$R_{a,N}$ on the direct lattice.}
\label{fig:zshodrtdist}
\end{figure}
Two conclusions stand out. First, $d_{h}$ and $d_{H}$ are
equal - i.e. the generated surfaces look the same at small as well as at large
distances. This is a non-trivial observation which is not always satisfied,
for example in multi-critical branched polymers\cite{our,bp}.
Then, there exists two different predictions for the value of $d_{H}$. One
is obtained by using the diffusion equation in Liouville theory\cite{diffL}
which predicts 
\begin{equation}
d_{H} = 2\frac{\sqrt{25-c}+\sqrt{49-c}}{\sqrt{25-c}+\sqrt{1-c}} = 3.561...
\end{equation} 
whereas redefining the distance in the context of transfer matrix 
theory\cite{sft} gives
\begin{equation}
d_{H} = \frac{2}{|\gamma (c)| } = \frac{24}{1-c+\sqrt{(25-c)(1-c)}} = 2
\end{equation} 
where $\gamma (c)$ is the string susceptibility. 
Comparing these with the results in table 1, it is apparent that the 
Liouville-prediction is strongly favored, whereas the prediction
from transfer matrix theory is disproved.

\end{document}